\documentclass[10pt,conference]{IEEEtran}
\usepackage[square, comma, numbers]{natbib}
\IEEEoverridecommandlockouts
\usepackage{comment}
\usepackage{tabularx}
\usepackage{makecell}
\usepackage{threeparttable}
\usepackage{amsmath}
\usepackage{amsfonts}
\usepackage{amssymb}
\usepackage{mathrsfs}
\usepackage[pdftex]{graphicx}
\usepackage{multirow}
\usepackage{color, colortbl}
\usepackage{enumitem}
\usepackage[dvipsnames]{xcolor}
\usepackage{pgfplots}
\usepackage{tikz}
\usepackage{tikzscale}
\usepackage{soul}
\usepackage{enumitem}
\usepackage{etoolbox}
\usepackage{xcolor}
\usepackage{tabulary}
\usepackage{subcaption} 
\usepackage[colorinlistoftodos]{todonotes}
\usepackage{algorithm, algorithmic}
\usepackage[normalem]{ulem}
\usepackage{caption}
\usepackage[export]{adjustbox}
\usepackage{fancyhdr}
\newcommand{\subparagraph}{}
\usepackage[compact]{titlesec}
\titlespacing{\section}{1pt}{*1.25}{*1.25}
\titlespacing{\subsection}{0.6pt}{*0.6}{*0.6}

\pgfplotsset{
  every axis plot/.append style={line width=1.2pt},
}

\newcommand{\cf}{\textit{cf.}, }
\newcommand{\eg}{\textit{e.g.}, }
\newcommand{\ie}{\textit{i.e.}, }

\newcommand{\fig}[1]{Fig.~\ref{#1}}
\newcommand{\tab}[1]{Table~\ref{#1}}

\newcommand{\sect}[1]{Section~\ref{#1}}
\newcommand{\red}[1]{\textcolor{red}{#1}}

\newcommand{\blue}[1]{\textcolor{black}{#1}}
\newcommand{\inlinemaketitle}{{\let\newpage\relax\maketitle}}

\pgfplotsset{compat=1.14}
\usepackage{newtxmath}
\usepackage{bbm}

\usetikzlibrary{arrows.meta, fit, backgrounds, shapes.geometric}
\tikzset{%
  >={Latex[width=2mm,length=2mm]},
         base/.style = {rectangle, rounded corners, draw=black,
                           minimum width=3cm, minimum height=1cm,
                           text centered, font=\rmfamily},
         terminal/.style = {base, circle, minimum size=1.5cm,font=\rmfamily}
}

\IEEEoverridecommandlockouts

\begin{document}
\bstctlcite{IEEEexample:BSTcontrol}

\title{Generalizable Resource Scaling of 5G Slices using Constrained Reinforcement Learning}
\author{
	\IEEEauthorblockN{
 	Muhammad Sulaiman\IEEEauthorrefmark{1}, Mahdieh Ahmadi\IEEEauthorrefmark{1}, Mohammad A. Salahuddin\IEEEauthorrefmark{1}, Raouf Boutaba\IEEEauthorrefmark{1}, and Aladdin Saleh\IEEEauthorrefmark{2}
 	}
 	\IEEEauthorblockA{
 	\IEEEauthorrefmark{1}David R. Cheriton School of Computer Science, University of Waterloo, Ontario, Canada\\
 	\texttt{\{m4sulaim,  mahdieh.ahmadi,  mohammad.salahuddin, rboutaba\}@uwaterloo.ca}
 	}
 	
  	\IEEEauthorblockA{
 	\IEEEauthorrefmark{2}Rogers Communications Inc., Ontario, Canada\\
 	\texttt{\{aladdin.saleh@rci.rogers.com\}}
 	}
}

\maketitle
\thispagestyle{fancy}
\IEEEpubidadjcol

\begin{abstract}
Network slicing is a key enabler for 5G to support various applications.   
Slices requested by service providers (SPs) have heterogeneous quality of service (QoS) requirements, such as latency, throughput, and jitter. It is imperative that the 5G infrastructure provider (InP) allocates the right amount of resources depending on the slice's traffic, such that the specified QoS levels are maintained during the slice's lifetime while maximizing resource efficiency.  However, there is a non-trivial relationship between the QoS and resource allocation. In this paper, this relationship is learned  using a regression-based model. We also leverage a risk-constrained reinforcement learning agent that is trained offline using this model and domain randomization for dynamically scaling slice resources while maintaining the desired QoS level. 
Our novel approach reduces the effects of network
modeling errors since it is model-free and does not require QoS metrics to be mathematically formulated in terms of traffic. In addition, it provides robustness against uncertain network conditions, generalizes to different real-world traffic patterns, and caters to various QoS metrics.
The results show that the state-of-the-art approaches can lead to QoS degradation as high as 44.5\% when tested on previously unseen traffic. On the other hand, our approach maintains the QoS degradation below a preset 10\% threshold on such traffic, while minimizing the allocated resources. Additionally, we demonstrate that the proposed approach is  robust against varying network conditions and inaccurate traffic predictions.

\end{abstract}

\section{Introduction}\label{sec:introduction}


With 5G, mobile networks are moving away from one-size-fits-all towards a more programmable network architecture. The adoption of Software Defined Networking (SDN) and Network Function Virtualization (NFV) allows an infrastructure provider (InP) to virtualize its physical network resources, and use them 
to create virtual isolated networks on top of a shared 
physical network infrastructure. These on-demand virtual isolated networks are also referred to as network slices.   Network slicing enables 
5G mobile networks to host applications or services with diverse quality of service (QoS) requirements. For example,  enhanced mobile broadband (eMBB) slices can be used for applications that require high throughput but lenient latency constraints such as 4K video streaming. On the other hand, ultra-reliable low-latency communication (URLLC) slices can be used for applications that require high reliability and very low latency such as remote surgery.

Whenever a service provider (SP) requests a slice from an infrastructure provider (InP), it includes its peak traffic and its required minimum quality of service (QoS) in the service level agreement (SLA). 
The required resources for maintaining a slice's QoS depend on the slice type and its traffic, which varies with time. The InP can guarantee the QoS  by allocating isolated resources to the slice based on its peak traffic. However, this can lead to over-provisioning since the actual traffic of a slice rarely reaches its peak \cite{traffic_dataset}. In this case, the majority of the allocated resources remain unused or under-utilized. 
On the other hand, the InP can improve its resource efficiency (RE) by predicting the future traffic of a slice, and preemptively scaling its resources accordingly. However, under-provisioning the resources, based on inaccurate traffic prediction or imprecise modeling of the relationship between allocated resources and QoS, can lead to a deterioration in the QoS of the slice. As a result, a certain level of QoS degradation is typically incorporated into SLAs, and the goal of InP is to dynamically scale resources to maximize resource efficiency while keeping QoS degradation under the specified limit. 
We refer to this as dynamic resource scaling. 

Several challenges need to be addressed to achieve effective dynamic resource scaling. Mobile networks consist of multiple domains including the radio access network (RAN), transport and the core network, and the QoS obtained can be dependent on the relative level of resources allocated in these domains. The QoS achieved at any resource allocation also depends on the state of the network at that time, \eg the level of interference by other slices or the state of the queues in the network. Additionally, the QoS may be defined heterogeneously for different slices serving different kinds of applications. For example, the QoS may be defined in terms of the throughput for an eMBB slice, and in terms of the latency for a URLLC slice.
Finally, even though some  proposed solutions in the literature require a dataset to be trained (e.g., \cite{icnp_older, LiWeighted}), the traffic that a slice experiences may be unknown during training time. 
Therefore, given the uncertainty of network conditions and future traffic, and the complex modeling of the end-to-end network, it is challenging to design an algorithm that can dynamically scale the resources of the slices while keeping their QoS degradation under the agreed-upon threshold.   

Several works in the literature model the mobile network mathematically \cite{li2020dra, overbooking}. Since it is quite challenging to accurately model end-to-end network dynamics \cite{inaccurate_modeling}, these works are based on simplifying assumptions which makes them inapplicable to real-world networks. 
Another method commonly used in the literature is to model the network as a queue and then use simulation to calculate the metrics of interest \cite{icnp_older, LiWeighted}. This method can be too simple to model a multi-domain mobile network,  
and complex user-level QoS metrics \cite{inaccurate_modeling, conext_paper}.
Given a network model, for resource scaling under QoS degradation constraints, a number of works in the literature propose using traditional deep reinforcement learning (DRL) algorithms such as Deep Q-Learning \cite{LiGAN, LiWeighted}. 
However, the reward function of these methods needs to be carefully engineered to achieve the desired tradeoff between QoS degradation level and resource efficiency, which is non-trivial.
To circumvent these problems, authors in \cite{liu2021onslicing} use constrained deep reinforcement learning (CDRL) and online learning to dynamically scale the resources. But online learning can be quite slow, and even infeasible for slices with short lifespans.
 

In this paper, we propose using a regression-based model to capture the behavior of an end-to-end network under different conditions. This model is trained offline using a dataset gathered by measuring the performance of an isolated slice in the real network under diverse network conditions and different amounts of allocated resources. Regression-based models, such as neural networks, can learn complex relationships between predictor and response variables, without requiring an exact network model.
For dynamically scaling the resources allocated to the slice while satisfying QoS requirements, we propose using CDRL with offline training. Although offline training addresses the slow training problem of online training, it must be generalizable to online traffic patterns not seen during offline training. For this purpose, we utilize a risk-constrained DRL algorithm coupled with domain randomization (DR). Risk-constrained DRL increases the chances of meeting QoS degradation constraints under unpredictable traffic and network conditions by constraining the risk rather than just the expected value of QoS degradation \cite{wcsac_paper}.  
DR is also a common technique for bridging the simulation-to-reality gap by randomizing the environment parameters during training  \cite{domain_rand}.
In addition, this RL agent is fed  with the output of an external traffic prediction module to avoid overfitting to any specific traffic pattern. Although the current evaluation is confined to the radio resource scaling and a single slice, it demonstrates the efficacy of the proposed solution. The contributions of this work are:

\begin{itemize}[leftmargin=*]
    
    \item We develop a novel framework for dynamic resource scaling which consists of a regression-based network model, risk-constrained DRL agent, and a traffic prediction module. By training the RL agent offline using random traffic, we have a generalizable agent that does not require any prior knowledge of online slice traffic patterns.  
    
    \item We evaluate the effectiveness of the proposed approach against traditional, and 
    constrained DRL-based models which encompass the state-of-the-art. In general, the proposed approach performs relatively better than others  while also showing generalization to previously unseen traffic and network conditions.
    \item We assess the robustness of the proposed approach under varying network conditions (\eg queue congestion), and inaccurate traffic predictions and demonstrate that it can effectively scale resources even under worse-case scenarios. 
    \item We show that our pre-trained model can be fine-tuned for increased performance while meeting QoS degradation constraint and  maintaining its generalization capability.

\end{itemize}

\blue{The rest of the paper is organized as follows. In \sect{sec:bg_related}, we provide an overview of CDRL, and discuss the related works. This is followed by a formal definition of the dynamic resource scaling problem in \sect{sec:statement}. \sect{sec:solutions} delineates the proposed solution. Finally, after showcasing the results in \sect{sec:experiments}, we conclude in \sect{sec:conclusion} and instigate future research directions.}

\section{Background and Related Works}\label{sec:bg_related}

\subsection{Constrained DRL---A Primer} \label{sec:contr_rl}
 In both traditional and constrained RL, the sequential decision making and interaction of an agent with its environment can be formally described using a Markov Decision Process (MDP) and a Constrained Markov Decision Process (CMDP), respectively. A finite-horizon MDP can be defined by the tuple ($O$, $A$,  $P$, $y$, $\rho_{0}$, $\gamma$), where $O$ is the state space, $A$ is the action space, $y: O \times A \rightarrow \mathbb{R}$ is the reward function, $P: O \times A \times O \rightarrow [0, 1]$ is the state transition probability distribution, $\rho_{0}$ is the initial state distribution, and $\gamma$ is the discount factor that specifies the relative importance of future rewards. 
 When state transition probabilities are unknown, RL can be adopted to find a policy $\pi : O \times A \rightarrow [0, 1]$ that can maximize the expected discounted reward defined as $J(\pi) =  \mathbb{E}_{(o_t, a_t)\sim\rho_\pi} \left[ \sum_{t} \gamma^t y(o_{t}, a_{t})\right]$, where $o_t$ and $a_t$ are the state and action at time step $t$, respectively, and $\rho_\pi$ denotes the state-action distribution induced by following policy $\pi$. 
 When the action and state spaces are large and/or continuous, the policy is learned using a parameterized deep neural network, which is known as DRL. 

CMDPs extend MDPs by adding cost functions $c: O \times A \rightarrow \mathbb{R}$ such that a CMDP is defined as ($O$, $A$, $P$, $y$, $c$, $\rho_{0}$, $\gamma$). Following a policy $\pi$, the cost distribution can be modeled as $p^\pi(c|s,a)$. The expected discounted cost $J_c(\pi)$ is analogous to $J(\pi)$ and is obtained by replacing the reward $y$ with cost $c$ in the corresponding equation, \ie $J_c(\pi) =  \mathbb{E}_{(o_t, a_t)\sim\rho_\pi} \left[ \sum_{t} \gamma^t c(o_{t}, a_{t})\right]$. In CMDPs, the objective is to find an optimal policy that maximizes $J(\pi)$, but also keeps $J_c(\pi)$ under a certain pre-defined threshold $c_{\textit{thresh}}$. Formally, this objective can be written as \cite{cpo_paper}: 
\begin{align}\label{eq:CDRL:obj}
    \pi^* =  & \quad \max_{\pi} J(\pi),\\
    \textrm{s.t.}& \quad J_c(\pi) \leq c_{\textit{thresh}}\notag. 
\end{align}

There are numerous approaches for solving CDRL problems (\cf \cite{CDRPsurvey}). 
In general, these methods are either based on Constraint Policy Optimization (CPO) \cite{cpo_paper}, or Lagrangian relaxation \cite{sac_lag, rew_constr}. CPO builds upon the TRPO algorithm \cite{schulman2015trust} by adding constraint satisfaction, leading to a monotonically improving policy that guarantees constraint satisfaction throughout training. On the other hand, Lagrangian relaxation methods work by relaxing constraints using Lagrangian multipliers, and updating decision variables and multipliers in an iterative manner using gradient ascent/descent. SAC-Lagrangian is one of these algorithms which is based on the actor-critic framework \cite{sac_lag}. It utilizes an actor network to represent the policy, and two critic networks to learn the expected reward and cost for any given state and action. These critic networks are then utilized to update the policy network 
to maximize the return while satisfying expected cost constraints. 


Although the methods discussed above satisfy a constraint on the expected value of the discounted cost distribution $p^\pi$, they do not constrain its variation. In this case, there remains a considerable probability of high-cost episodes. As a result, such methods can not be used in safety-critical applications, where it is crucial for the learned policies to be robust. 
For this case, rather than the expected cost, the safety metric of interest is defined in terms of the Conditional Value-at-Risk (CVaR) of the cost, \ie
\begin{equation}
    \textit{CVaR}_\alpha = \mathbbm{E}_{p^\pi} [C | C\geq F_C^{-1}(1-\alpha)],
\end{equation}
where $F_C(.)$ is the cumulative distribution function (CDF) of $p_\pi$, and $\alpha$ is the risk level hyper-parameter with smaller values leading to more risk-averse policies. In this case, a policy is safe if it satisfies a constraint on this new safety measure:
\begin{equation}
    \mathbbm{E}_{p^{\pi}} [C(o_t, a_t) | C(o_t, a_t) \geq F_C^{-1}(1-\alpha)] \leq c_{\textit{thresh}}, \;\;\;\;\forall t,
\end{equation}
where $C(o_t, a_t)$ is the discounted cumulative cost of policy $\pi$ from point $(o_t, a_t)$.

To address this issue, \citet{wcsac_paper} recently proposed the Worst-Case Soft Actor Critic (WCSAC) algorithm that extends SAC-Lagrangian by replacing the cost critic with a distributional one. Specifically, they model $p^\pi$ as a Gaussian distribution, and utilize two neural networks to predict its mean $Q^c_\pi(s_t,a_t)$ and variance $V^c_\pi(s_t,a_t)$ given a state-action pair $(s_t, a_t)$. The new safety measure, \ie CVaR, can then be calculated using a closed-form equation:
\begin{equation}
    \textit{CVaR}_\alpha=\Gamma_\pi(s_t,a_t,\alpha) = Q^c_\pi(s,a) + \alpha^{-1}\phi^{-1}(\Phi^{-1}(\alpha)),
\end{equation}
where $\phi(.)$ and $\Phi(.)$ represent the probability distribution function (PDF) and CDF of the standard normal distribution, respectively. The policy can be updated by minimizing the following KL-Divergence \cite{kl_div, wcsac_paper}:
\begin{equation}
    \pi' = \min_{\pi} \mathcal{D}_\textit{KL}\left(\pi(.|s_t) \bigg\vert\bigg\vert \frac{\textit{exp}(\frac{1}{\beta} Q^r_\pi(s_t,.))-k \Gamma_\pi(s_t,.,\alpha)}{Z^{\pi}(s_t)} \right)
\end{equation}
 where $\pi^{'}$ denotes the updated policy, $Q^r_\pi(s,a)$ denotes the state-action value and $Z^\pi(s_t)$ is a normalization factor. $\beta$ and $k$ are Lagrangian multipliers that are updated iteratively to determine the trade-off between the policy entropy, the reward, and the safety measure. For a detailed exposition of the algorithm, we refer to \cite{wcsac_paper}. 
As WCSAC is able to satisfy a percentile-based constraint (CVaR) over a QoS degradation distribution, compared to expectation-based methods, it is more adaptable to different traffic patterns and network condition scenarios.
 

\subsection{Dynamic Resource Scaling}\label{sec:related}
Dynamic resource scaling requires an accurate model of the relationship between the allocated resources, traffic volume, and achieved QoS. 
In this section, we present an overview of the methodologies commonly used to model this relationship and review state-of-the-art dynamic resource scaling works that fall into these categories.

\subsubsection{Resource Isolation}

In this approach, QoS is defined in terms of \textit{resource isolation} that depends on a simple comparison between the required resources and the resource allocation, \ie $\mathbb P(r^s_t \geq v^s_t)$, where $r^s_t$ and $v^s_t$ respectively denote the amount of allocated resource to a slice, and its  minimum  required resource to meet the QoS threshold at a specific time $t$. This type of network modeling requires ${v}^s_t$ to be known or easily predictable which may not be feasible in practice. 
Additionally, 
this may be undesirable for cases when the slices experience bursty traffic since even if only a small amount of resources are allocated to the slice during short periods of high traffic, high isolation can still be achieved.
Based on this model, \citet{li2020dra} optimize the resource utilization at each time step while respecting resource isolation constraints modeled by chance constraints. These constraints are then approximated using a data-driven approach, which converts the problem to  a semi-definite programming (SDP) problem that can be solved optimally.

\subsubsection{Model-Driven QoS Isolation}

This approach  adheres to \textit{QoS isolation} by calculating the minimum required resources based on a given QoS threshold, but only considers an abstract model of the network for which there exists concrete theoretical groundwork.
Queues are one of the commonly used models in this category. In this case, either queuing theory or queue simulation can be utilized to compute different performance measurements for the packets entering the queue.
By defining the queue arrival-rate in terms of slice traffic and determining the service-rate based on resource allocation, the queuing-time for any packet can be determined.
The average queuing-time can represent the QoS of a slice, defined as the packet latency. 
In general, the drawback of this type of modeling is that it can only deal with simple network-level QoS metrics which are based on a single resource type, as it can be challenging to model the service-rate of the queue based on multiple heterogeneous metrics.
Solutions based on queue simulation also suffer from high computational complexity when the traffic volume increases since they deal with the traffic at the packet-level. Finally, queues might not be able to accurately model network components such as RAN, as there can be additional factors that affect the latency such as the presence of multiple queues, channel effects or application-specific idiosyncrasies.

Some of the works that use a queue for modeling the network include \cite{papa2019optimizing, kasgari2018stochastic, icnp_older, LiWeighted, LiGAN}. \citet{papa2019optimizing} and \citet{kasgari2018stochastic} assume the delay experienced by each slice can be exactly modeled by considering an M/M/1 queue model. The authors then leverage the Lyapunov optimization method to minimize long-term resource usage under QoS constraints. 
\citet{LiWeighted} and \citet{LiGAN} utilize queue-based simulation and optimize the weighted sum of RE and SLA satisfaction ratio. For this purpose, \cite{LiWeighted} utilizes Deep Q-learning, whereas \cite{LiGAN} leverages a combination of distributional DRL with generative adversarial network (GAN). However, in these works, the trade-off between RE and SLA satisfaction ratio depends on the weights assigned to these in the reward formulation, which have to be manually tuned.
As a result, these approaches are not able to satisfy QoS degradation constraints. To ensure these constraints, the authors in \cite{icnp_older} propose constrained RL for resource allocation. But this approach requires the slice's traffic to be known in advance and 
is unable to generalize to previously unseen traffic patterns during testing.

\subsubsection{Data-Driven QoS Isolation} \label{sec:datadriven_model}
In this method, QoS isolation is assured and  the  behavior  of the network is learned using data-driven approaches based on historical data.
As there is no public dataset available, this mandates access to either a testbed or a production network where slices can be easily created and scaled. Due to this reason, there are only a limited number of works that utilize this approach. This type of network modeling is classified further into the following categories:

\begin{figure}[!t]
\centering
\includegraphics[width=\linewidth]{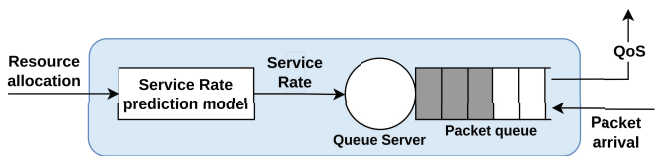}
\caption{Regression-Queue based network model}
\label{reg_queue_model}
\end{figure}

\textbf{Regression-Queue Based:} As shown in Fig. \ref{reg_queue_model}, this approach is also based on a queue, with the difference that a regression model, learned using a network dataset, is used to predict the service rate of the queue. Since multiple resources can be used as the input to the service rate prediction model, this type of network modeling is not restricted to only one resource type. 
However, this approach remains constrained by other drawbacks associated with queue-based model-driven approaches.
\citet{conext_paper} used offline training based on this model to minimize multi-domain resource utilization under capacity and end-to-end delay constraints of slices. They relax SLA satisfaction constraints and incorporate them into the objective by utilizing the primal-dual Lagrangian method, and train the model using conventional actor-critic DRL methods. Although offline training reduces the time for the RL model to converge in the real environment, similar to \cite{icnp_older} it suffers from low generalizability. 

\textbf{Online Learning Based:
}In this approach, rather than modeling a network, the data from a production network during a slice's operation is leveraged to learn the resource scaling algorithm.
In a subsequent work to \cite{conext_paper}, \citet{liu2021onslicing} proposed  an RL-based approach with online-only training and behaviour cloning, which can deal with various QoS metrics. However, considering the granularity of updates in a real network (\ie around 15 minutes \cite{liu2021onslicing}), the adoption of this method in a production environment is impractical due to the long convergence time.

\section{Problem Statement}\label{sec:statement}
Let $TTI$ be the transmission time interval, \ie small isometric time intervals into which the time horizon is divided. We define $T_{\Delta}$, consisting of $N$ TTIs, to be the decision time interval (DTI), \ie the minimum time interval required between resource scaling decisions due to practical limitations such as the time required for horizontally or vertically scaling virtual machines (VMs). Let $\mathbf{T}$ denote the set of starting points of all DTIs. For slice $s \in S$, we denote the traffic at the DTI starting at $t$ by column vector $\mathbf{x}^s_{t}=[x_{tn}^s]_{n \in [N]}$, where $x_{tn}^s \in \mathbb{N}$ is the traffic at the $n^{\text{th}}$ TTI within DTI starting at $t$. We also define  $\mathbf{r}^s_t = [r^{s,k}_t]_{k \in [K]}$ with $r^{s,k}_t \in \mathbb{R}$ to be the $K$ different types of  resources allocated to slice $s$ over DTI $t$. We assume that the traffic, measured in users/sec, stays constant within a TTI and the resources allocated to a slice are divided fairly among all its users.

Let vector $\mathbf{q}_t^s=[q_{tn}^s] \in \mathbb{R}^{n \times 1}$ represent per-user QoS of each slice  at DTI $t$, which can be determined using a network model. We are interested in QoS degradation probability at any point in time, which can be defined as the portion of traffic that receives QoS below the minimum threshold, \ie
\begin{equation} \label{beta}
\begin{aligned}
\beta^s_t = \frac{\sum_{\substack{\tau \in T: \tau \leq t}}  {\mathbf{x}^s_\tau}^\intercal \, \blue{\mathbbm{1}_{[q^s_{\tau} \leq q^s_{\textit{thresh}}]}}}{\sum_{\tau \in T: \tau \leq t} {\mathbf{1}_N \, \mathbf{x}^s_\tau}},
\end{aligned}
\end{equation}
where $\mathbbm{1}_{[q^s_{\tau} \leq q^s_{\textit{thresh}}]}$ is an indicator vector whose $n^{\text{th}}$ element equals to $1$ only when $q^s_{\tau n} \leq q^s_{\textit{thresh}}$, $\mathbf{1}_N$ is a 1-vector of size $N$ and $q^s_\textit{thresh}$ is the expected minimum QoS of slice users. Finally, based on the introduced notations, the dynamic resource scaling problem can be formulated as:
\begin{equation} \label{obj}
\begin{aligned}
\min_{\mathbf{r}} \quad & \dfrac{1}{|T|} \sum_{t \in T} \sum_{s \in \mathbf{S}} \eta^\intercal  \mathbf{r}^s_t\,\\
\textrm{s.t.} \quad & \mathbbm{E} \left( \beta^s_{\textit{max}(T)}\right) \leq \beta_{s, \textit{thresh}}, \;\;\;\forall s \in S\,\\
\quad & \sum_{s \in S} \mathbf{r}_t^s \leq {\mathbf{R}}, \;\;\; \forall t \in T,\\
\end{aligned}
\end{equation}
where $\eta \in \mathbb{R}^{k \times 1}$ is the resource normalization vector, $\mathbbm{E}$ is the expectation over the distributions of QoS and traffic, $\mathbf{R}=[R_k]_{k \in [K]}$ represents the capacities of resources, and $\beta_{s, \textit{thresh}}$ is the acceptable $QoS$ degradation threshold for slice $s$.

\section{Proposed Framework}\label{sec:solutions}
In this section, we describe the proposed framework which is shown in Fig. \ref{train_loop}. It consists of three components: a future traffic forecast module, an RL-based dynamic resource scaler and a network model. Note that we do not expound on the traffic prediction module, since it is a well-studied topic and there are off-the-shelf packages available for it (\eg \cite{neural_prophet}).

\subsection{Network Model} \label{sec:reg_model}
We propose using a regression-based network model. Similar to the state-of-the-art  \cite{LiGAN, liu2021onslicing}, we assume that there are only a limited number of standard slice types (\eg URLLC, eMBB, mMTC), and their QoS can be monitored by creating isolated slices and scaling their resources. Compared to the regression-queue based model in \sect{sec:datadriven_model}, the queue and the service rate prediction model are replaced by a single regression model. This regression model learns the function $f(x, \mathbf{r}): \mathbb{N} \times \mathbb{R}^{K} \rightarrow \mathbb{R} \times \mathbb{R}$, which maps the traffic $x^{s}_{tn}$ and resource allocation vector $\mathbf{r}^s_t$ of $n^{\text{th}}$ TTI of DTI $t$  to Gaussian distribution parameters $\mu_{tn}^s$ and $\sigma_{tn}^s$. The QoS can then be sampled using these parameters, \ie $q_{tn}^s \sim \mathcal N(\mu, \sigma)$. In this method, we assume that the effect of past traffic on the QoS is only transient, and can be subsumed by the distribution of QoS around the mean. This is because it can be infeasible to gather a dataset that correlates the past traffic as well as the current traffic, and a resource allocation to a certain QoS. 

Based on the complexity of the network (\eg in terms of the number of resources), a simple query-based method, a linear regression model, or neural networks could be used to learn the function $f(x, \mathbf{r})$. To train this model, the dataset is gathered by performing a grid search over different resource allocations at different traffic levels, and measuring the corresponding QoS under varying network conditions. As each QoS value sampled from the Gaussian distribution $\mathcal N(\mu, \sigma)$ can be written as $\mu - d\sigma$, we use the parameter $d \in \mathbb{R}$ to represent network conditions such as queue congestion and channel quality. When $d$ is known, \ie in deterministic network conditions, the QoS vector at DTI $t$ can be computed as:
 \begin{equation}
\begin{aligned} \label{eq:det_cond}
\textit{q}^{s, \textit{det}}_t(d) = \left[\mu_{tn}^s - d\sigma_{tn}^s\right]_{ n \in [N]}.
\end{aligned}
 \end{equation}


As opposed to queue-based models, the regression-based model can deal with scaling heterogeneous resources, and can predict different types of QoS. Additionally, the complexity of this approach does not depend on the traffic volume and it is not restricted to packet-level traffic.



\subsection{Risk-constrained DRL-based Resource Scaling Algorithm} \label{sec:design}
\begin{figure}[!t]
\centering
\includegraphics[width=\linewidth]{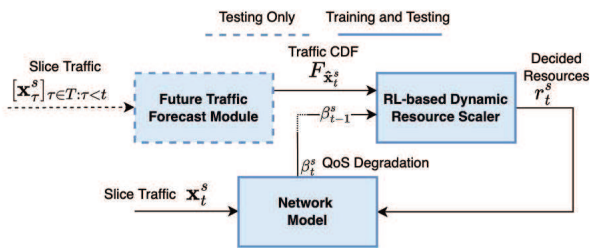}
\caption{Overview of the proposed framework}
\label{train_loop}
\end{figure}

For dynamically scaling the resources allocated to a slice, we propose using the WCSAC \cite{wcsac_paper} algorithm trained using the training loop shown in Fig. \ref{train_loop}. Each step in an episode, \ie $t \in T$, corresponds to a DTI.  
\subsubsection{CMDP Formulation}
At step $t$, the RL agent's \textit{state} $o_t$ includes the CDF of the traffic of the next DTI over its TTIs, $F_{x^s_{tn}}(.)$, and the QoS degradation up until that time, $\beta^s_{t-1}$, for each slice. Note that since the traffic is in terms of users/sec,  $F_{x^s_{tn}}(a)=\mathbb{P}(x^s_{tn} \leq a)$ is a discrete CDF which can be represented by a vector of finite size. As opposed to existing works  \cite{conext_paper, icnp_older}, instead of mean, we include the CDF of the future traffic in the agent's state space. This is because multiple different distributions \blue{can have the same mean, and can lead the agent to learn spurious correlations between actions in a given state and their effect on the environment \cite{confounding_vars}}. Additionally, $\beta^s_{t-1}$, which is calculated using  Eq.~\eqref{beta}, acts as a feedback mechanism for the agent to adjust the resource allocation for the future DTIs based on the effect of past actions on QoS degradation. 
The agent's \textit{action} $a_t$ corresponds to the resource scaling decision $\mathbf{r}_t^s$ for each slice and 
we define the \textit{reward} function as:
\begin{equation} 
\label{eqn:wcsac_rew}
y(o_t, a_t) = 1 - \sum_{s \in \mathbf{S}} \eta^\intercal \mathbf{r}^s_t.
\end{equation}
Finally, the cost function is defined as the marginal QoS degradation where the denominator is the sum of traffic over the episode, as shown below: 
\begin{equation}
\label{eqn:wcsac_cost} 
c(o_t, a_t) = \frac{{\mathbf{x}^s_t}^\intercal \  \mathbbm{1}_{[q^s_{t} \leq q^s_{\textit{thresh}}]}} { \sum_{\tau \leq max(T)} \mathbf{1}_N \, \mathbf{x}^s_\tau}.
\end{equation}

\subsubsection{Training}

\fig{train_loop} shows how WCSAC algorithm is integrated with the network model and future traffic prediction module during training and testing. To make the learned policy generalizable across different traffic patterns and network conditions, we utilize uniform DR during training. For this purpose, at the start of each episode, the traffic for each DTI, $\mathbf{x}^s_t$, is generated by sampling i.i.d from a randomized distribution, \ie $F_{x^s_{tn}}$. By randomizing the traffic distribution and sampling the QoS from the network model distribution during training, the risk-constrained RL agent learns to maximize the reward while keeping the QoS degradation under the specified threshold $\beta^s_{\textit{thresh}}$ even in the  worst-case scenarios. These extreme conditions can arise due to congestion in the network, \blue{interference by other slices}, or a specific traffic pattern (\eg bursty) that may require higher resource allocation. 
 
Since the future traffic, $\hat{\mathbf{x}}^s_t$, is unknown during testing, an external traffic prediction module is used to predict its CDF. The QoS degradation during testing can be computed using actual slice traffic, and by utilizing either the offline network model or actual QoS reported by users. Note that by including the future traffic distribution in the state space, the agent is not confounded by the varying effect of different resource allocations under varying traffic levels.
 


\section{Experiments}\label{sec:experiments}

\subsection{Testbed and Simulation Setup}
To gather the required dataset for creating a realistic network model, we deploy the SDRAN-in-a-Box (RiaB) \cite{riab} on an Intel NUC PC. RiaB utilizes Kubernetes to deploy end-to-end SD-RAN components that include EPC, emulated RAN and user equipment (UE), and ONOS RAN Intelligence Controller (RIC). The testbed can be used to create slices, associate UEs to them, and scale their allocated resources dynamically. However, there are limitations on the number of UEs per slice and the granularity of resource allocation. To circumvent the former restriction, we count the number of parallel connections made by an emulated UE as the slice's users. For the latter, we allocate radio resources in intervals of $10\%$ of the total resource capacity. For simplicity, during evaluation, we consider a single type of slice (\ie eMBB) and resource (\ie radio resource in Mbps), and thus we omit the subscript $s$ and $k$ in the subsequent sections. The slice is tailored for eMBB services by deploying an image download application over it. We define the $\textit{QoS}$ of the slice in terms of frames per second (Fps), \ie the number of times the image can be downloaded from a local server within a second. 

\begin{figure}[!t]
	\centering
        \includegraphics[width=0.735\columnwidth]{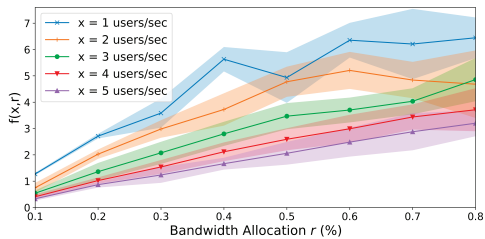}
        \caption{Regression-based network model}
        \label{network_model}
\end{figure}

To gather the QoS dataset, we perform a grid-search over the alterable parameters by varying the traffic from 1 to 5 users/sec, and the corresponding resource allocation from $10\%$ to $80\%$. To encompass different network conditions, we repeat this grid-search multiple times and gather the data for 30 seconds for each point in the grid per iteration. \fig{network_model} shows the corresponding network model. The shaded area around the mean represents the distribution of QoS achieved at different network conditions. The network model is fed with this data to predict the distribution of $QoS$ at any traffic and resource allocation.
In this paper, we learn the network model using a simple query-based model that returns the mean and standard deviation of the QoS as the size of the grid is small. 
For training and testing the different scaling approaches, we set the QoS threshold and the acceptable QoS degradation threshold, \ie $q_{\textit{thresh}}$ and $\beta_{\textit{thresh}}$, to $2.0$ Fps and 10\%, respectively. As we can see in \fig{network_model}, $80$\% bandwidth allocation is required to have no (\ie 0) QoS degradation under the highest traffic rate (\ie 5 users/sec) and worse-case network conditions which occur at around $d=-2$.
For WCSAC, we set the risk level hyper-parameter $\alpha$ to 0.1. DTI and TTI are set to one minute and one second, respectively. As discussed previously, we assume the traffic varies across TTIs, but remains constant within each TTI. For testing, we utilize the real-world traffic pattern of Internet events in Milan from the Telecom Italia dataset \cite{traffic_dataset}. We scale this traffic pattern to represent the traffic of 1 to 3 users/sec and add truncated Gaussian noise $\mathcal N(0, 0.75^2)$ to create dynamic traffic within each DTI. We refer to the resulting traffic curve as the \textit{dataset traffic} throughout this section. The reported results are averaged over 100 episodes of 10 minutes in length to ensure statistically stable results.


 \subsection{Comparative Approaches}
 
 For evaluating the proposed framework comprising the regression-based network model, WCSAC algorithm and described training paradigm, we implement a number of DRL-based approaches, and a heuristic method. These  approaches encompass the baselines and emulate a number of the solutions proposed in contemporary literature. For training and testing these methods, we utilize the proposed regression-based network model for a fair comparison.

 \subsubsection{DRL-based}
 \textbf{\small{Avg-CPO},\;\small{Avg-PPO}: }A number of DRL-based works in the literature train the RL agent on the same traffic pattern which is also used during testing \cite{icnp_older, conext_paper, LiGAN, LiWeighted}. Clearly, these approaches fail to perform well if the test traffic varies from the training traffic.
 To encompass both constraint-aware and traditional DRL approaches presented in these works, we implement the CPO \cite{cpo_paper} and the PPO \cite{ppo} algorithms and refer to them as {\small{Avg-CPO}} and {\small{Avg-PPO}}, respectively.
 For {\small{Avg-CPO}}, the reward and  cost are the same as the ones used for WCSAC, described in equations \eqref{eqn:wcsac_rew} and \eqref{eqn:wcsac_cost}. Since a cost constraint cannot be incorporated with PPO, we formulated its reward function as the sum of the reward and the cost  used for WCSAC, weighted by $w_{\textit{RE}}$ and $w_{\textit{QoS}}$, respectively. The values for these weights are manually fine-tuned. 
 Some of these works assume that the RL agent learns to scale the slice resource allocation only based on past traffic information \blue{\cite{conext_paper, icnp_older}}. This requires the RL agent to predict the traffic trend in addition to resource scaling. However, for a fair comparison, 
 we include the CDF of the future traffic in the RL agent's state. 

 \textbf{{\small WC-CPO}: } As described in \sect{sec:contr_rl}, CPO is designed to constrain the expected cost under a specific limit while learning actions that maximize the reward. To make this approach risk-aware, \ie robust to different traffic distributions and network conditions that arise during testing, we leverage cost-shaping to ensure that the QoS degradation stays below  $\beta_{\textit{thresh}}$ for all scenarios. For this purpose, in addition to the per-step cost in Eq.~\eqref{eqn:wcsac_cost}, the agent is given an additional cost at the end of each episode. This cost is an exponential function of QoS degradation that is in excess of the threshold, \ie $\gamma(e^{[ \beta_{\textit{max}(T)} - \beta_{\textit{thresh}}]^+} - 1)$ where $[x]^+$ denotes the maximum of $x$ and 0, and $\gamma$ is a scaling factor which controls the degree of risk adverseness. Since the average cost is affected significantly by this exponentially weighted cost component, the agent learns to keep the QoS degradation for scenarios giving rise to high QoS degradation under the threshold. We refer to this CPO-based approach with the shaped cost function as worst-case CPO, and denote it by  {\small WC-CPO}.
 

\begin{figure*}[t!]
	\centering
	\begin{subfigure}[c]{0.7\columnwidth}
	\centering
    \includegraphics[width=\linewidth]{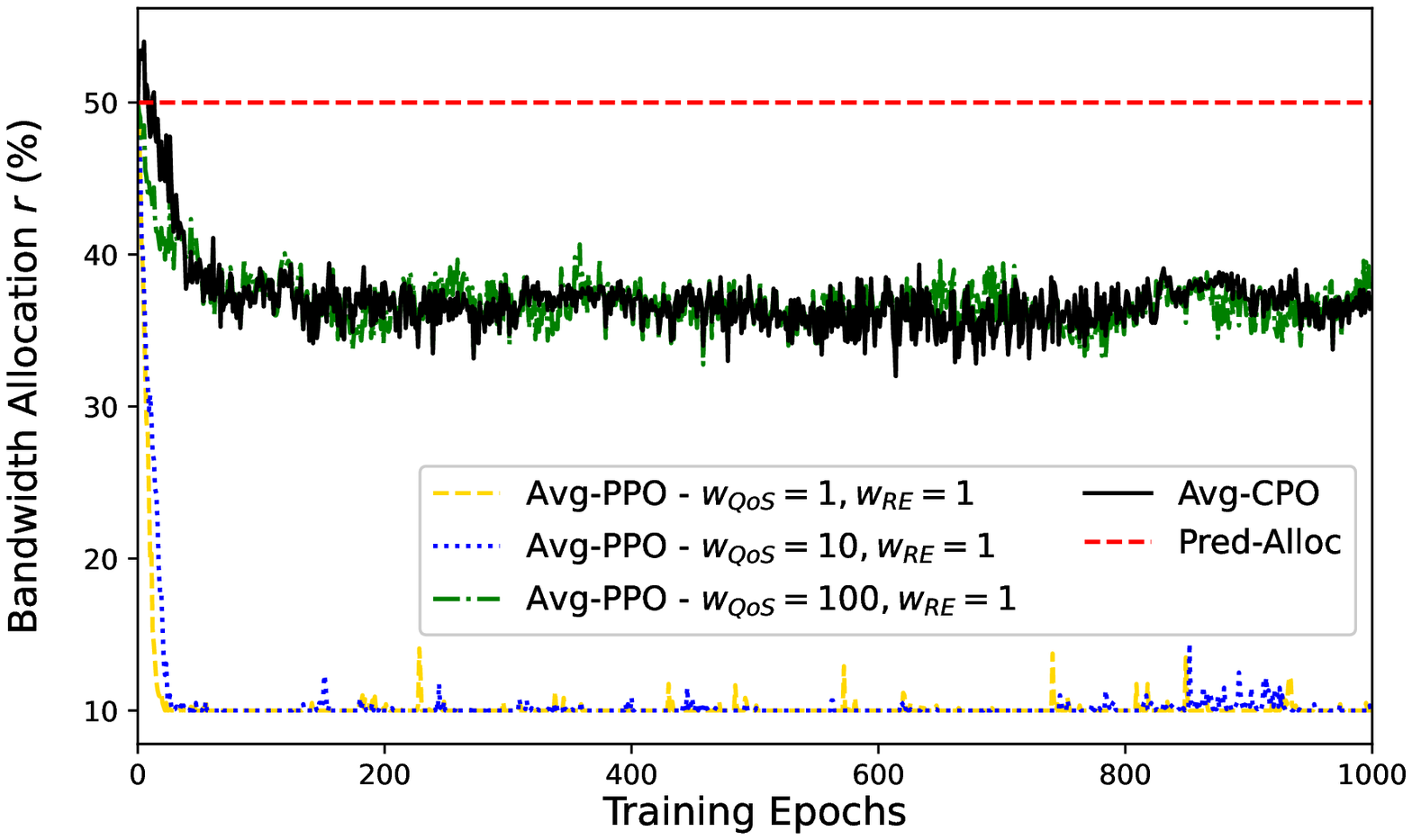}
    \caption{Avg. BW allocation per epoch }
    \label{trad_dw}
    \end{subfigure}\qquad\qquad\qquad
    \begin{subfigure}[c]{0.7\columnwidth}
	\centering
    \includegraphics[width=\linewidth]{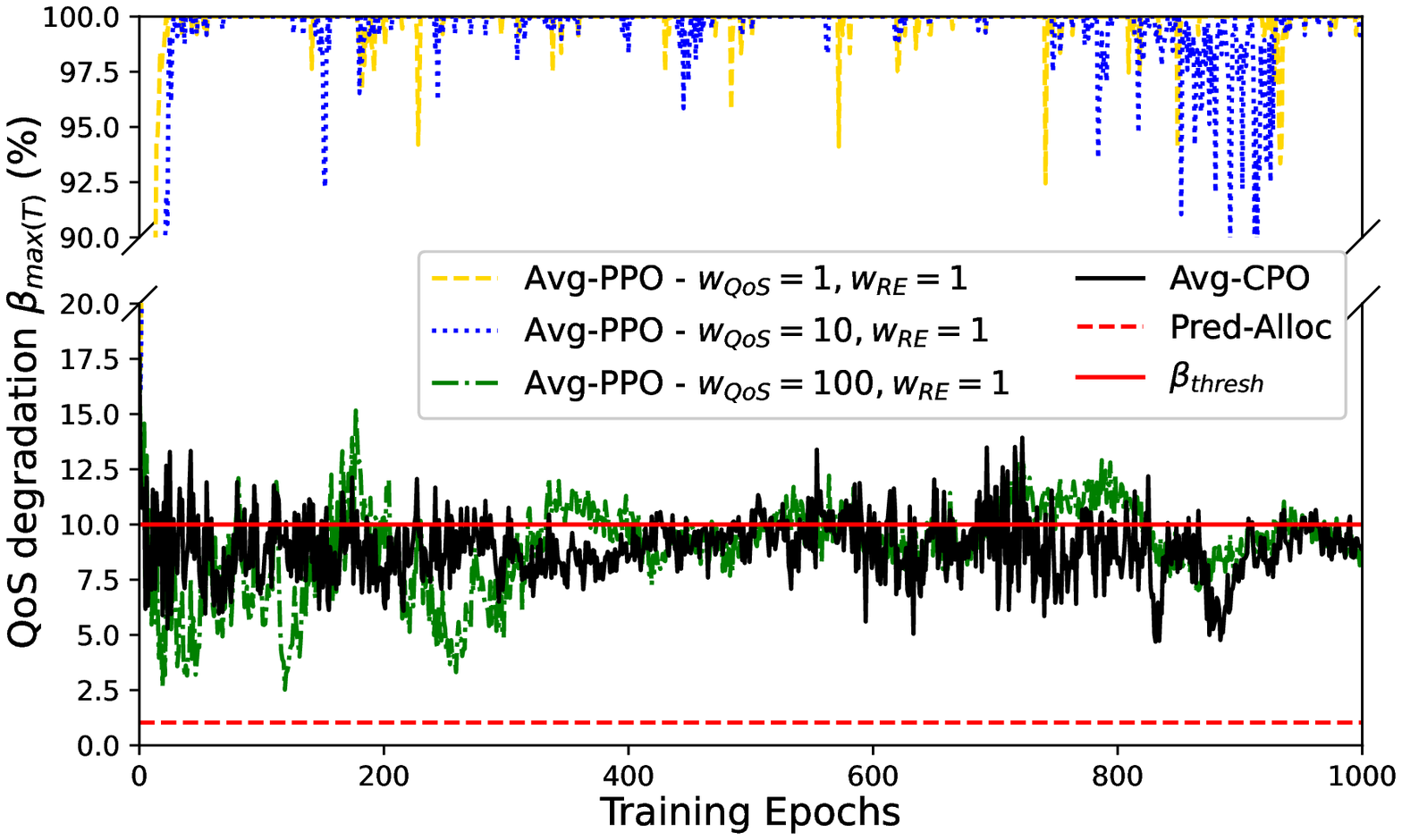}
    \caption{Avg. QoS degradation per epoch ($\beta_\textit{thresh}=10\%$)}
    \label{trad_sla}
\end{subfigure}
\begin{subfigure}[t]{0.7\columnwidth}
\centering
\includegraphics[width=\linewidth]{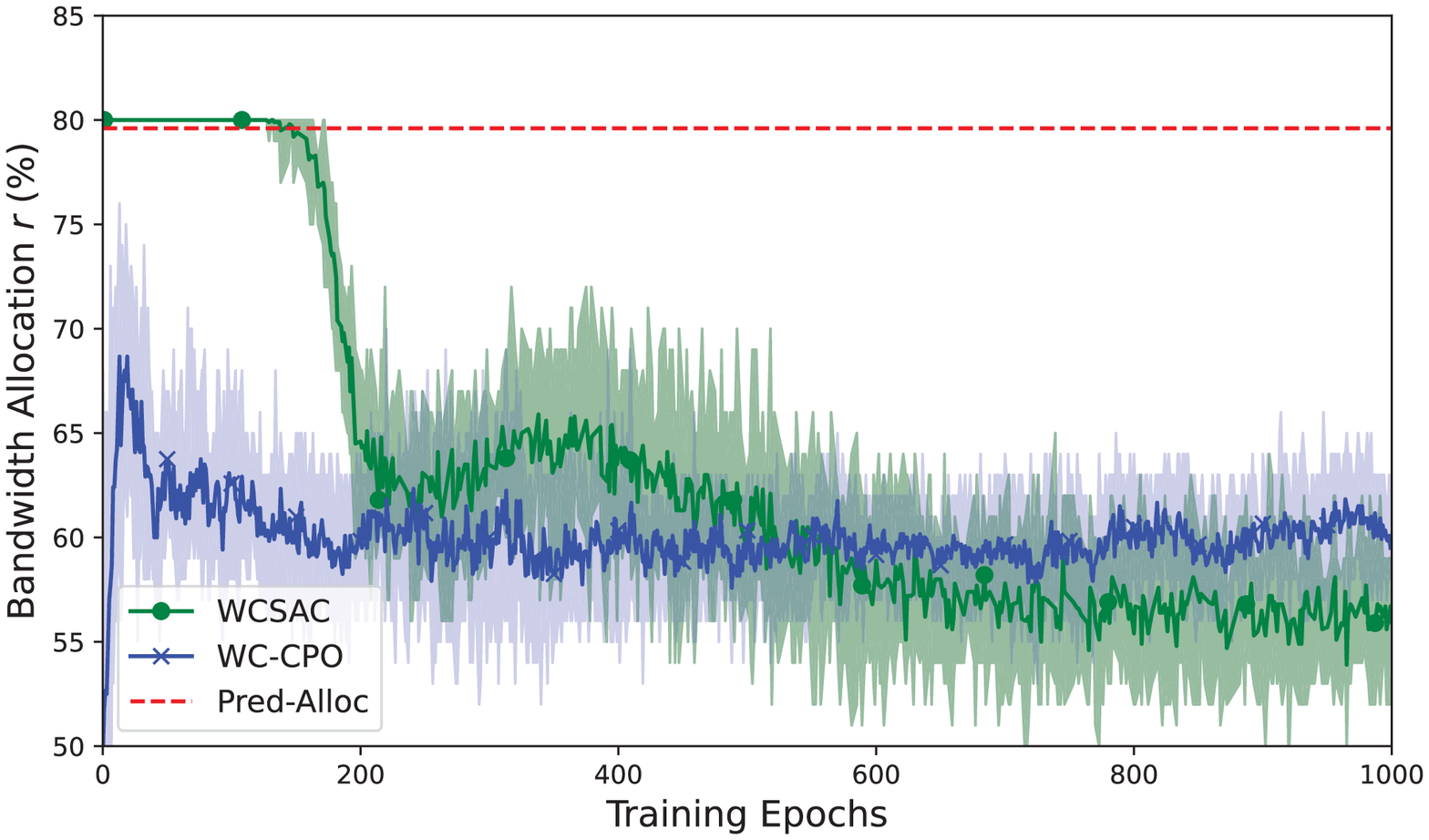}
\caption{Avg. BW allocation per epoch }
\label{wc_dw}
\end{subfigure}\qquad\qquad\qquad
\begin{subfigure}[t]{0.7\columnwidth}
\centering
\includegraphics[width=\linewidth]{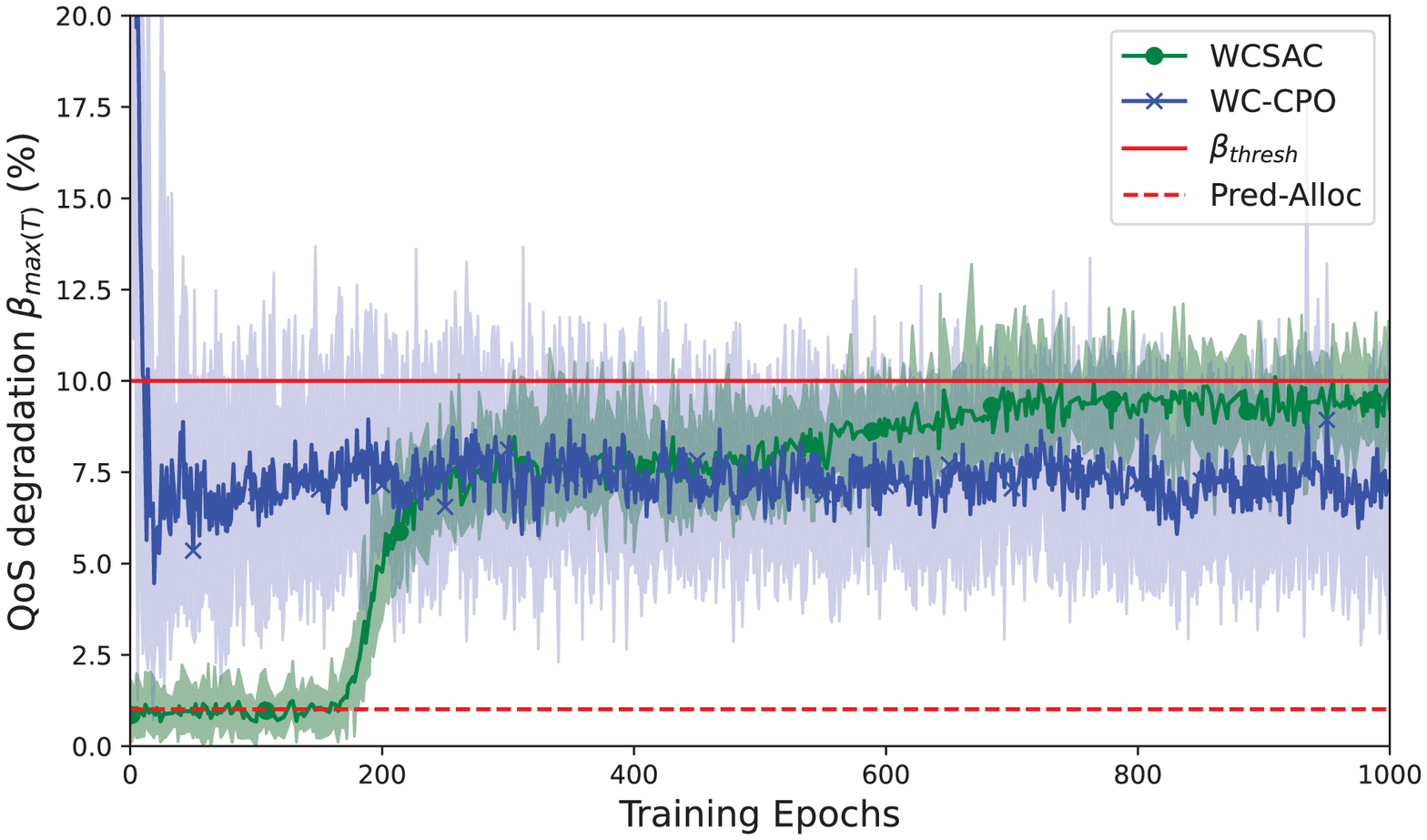}
\caption{Avg. QoS degradation per epoch  ($\beta_\textit{thresh}=10\%$)}
\label{wc_sla}
\end{subfigure}
\caption{Training process of different methods; (\subref{trad_dw}),(\subref{trad_sla}) are trained on the dataset traffic, and (\subref{wc_dw}),(\subref{wc_sla}) are trained on randomized traffic.}
\end{figure*}
 
 \subsubsection{Heuristics-based}
 \textbf{{\small Pred-Alloc}:}  As an upper bound for evaluating DRL-based approaches, we also implement a simple heuristic. This algorithm is based on two simplifying assumptions. First, it assumes the traffic in any DTI to be constant and equal to the peak of the curve predicted by the traffic prediction module. The second simplification is that it considers somewhat extreme and deterministic network conditions, \ie at any given traffic and resource allocation, the QoS vector is equal to $q^{det}_t(-2)$ defined in Eq.~\eqref{eq:det_cond}. These assumptions make the resource allocation at each DTI trivial since it can be calculated using a simple grid-search. However, it can lead to significant over-provisioning and non-optimal results. We refer to this approach as {\small{Pred-Alloc}}. 

\subsection{Training Performance}

For training, we use two dense layers each having 64 neurons for the policy network for the different algorithms. The learning rates can differ across methods and are determined experimentally. We train {\small{Avg-CPO}} and {\small{Avg-PPO}} on the dataset traffic. For {\small{Avg-PPO}}, we set $w_{RE}$ to 1, and tune $w_{QoS}$ to achieve a nearly optimized  tradeoff between resource allocation and QoS. 
\fig{trad_dw} and \fig{trad_sla} show the mean bandwidth allocation and mean QoS degradation of these methods during the training phase. We observe that {\small{Avg-PPO}} is able to keep QoS degradation under 10\% with $w_{QoS}$=100. 
When evaluated, on the same traffic distribution as the one used for training (\cf \tab{table:ev}), {\small{Avg-CPO}} and {\small{Avg-PPO}} ($w_{QoS}$=1, $w_{RE}$=100) achieve the same mean bandwidth allocation of 37.2\%, and a mean QoS degradation of 8.42\% and 8.3\%, respectively. This bandwidth allocation is 12.8\% lower than {\small{Pred-Alloc}}, which allocates a 50.8\% mean bandwidth in this scenario. 

\fig{wc_dw} and \fig{wc_sla} show the training progress of {\small{WC-CPO}} and {\small{WCSAC}} approaches on the randomized traffic. The shaded area around the curves denotes the minimum and maximum of the corresponding quantities. 
The variation range of mean bandwidth and QoS degradation represent the 
ability of the two approaches to operate under different network conditions and traffic distributions. 
As reported in \tab{table:ev}, when evaluated on dataset traffic, {\small{WC-CPO}} and {\small{WCSAC}} lead to a mean bandwidth allocation of 52.5\% and 39\%, and a mean QoS degradation of 1.06\% and 6.73\%, respectively. In this scenario, the respective approaches lead to 2.5\% higher, and 11.8\% lower bandwidth allocation compared to {\small{Pred-Alloc}}.

\subsection{Generalization to Unseen Traffic Pattern}

\begin{table}[!t]
\footnotesize
\caption{Evaluation Performance of Different Approaches}
\centering
\label{table:ev}
\resizebox{1.0\columnwidth}{!}{
\begin{tabular}{|l|c  c | c c|}
\hline
\multirow{2}{*}{\textbf{Method}}&\multicolumn{2}{c|}{\textbf{Dataset Traffic}}&\multicolumn{2}{c|}{\textbf{Offset Dataset Traffic}}\\
&BW allocation(\%)&QoS degradation(\%)&BW allocation(\%)&QoS degradation(\%)\\\hline
     Avg-PPO&37.2&8.3&39.7&40.8\\
     Avg-CPO&37.2&8.42&39.4&44.5\\
     WC-CPO&52.5&1.06&53.9&8.98\\
     WCSAC&39&6.73&56&7.19\\
     Pred-Alloc&50.8&1.01&77.7&0.995\\\hline
\end{tabular}}
\end{table}
To assess generalization, we offset the traffic curve by 2 users/sec, \ie the trained agents are evaluated on the \textit{dataset traffic} that represents 3 to 5 users/sec. The results are reported in \tab{table:ev}. In this scenario, {\small{Avg-CPO}} and {\small{Avg-PPO}} lead to similar mean bandwidth allocation of 39.4\% and 39.7\%, but a mean QoS degradation of 44.5\% and 40.8\%, respectively. The results assert that these methods over-fit to the training traffic pattern and are unable to generalize to previously unseen traffic, leading to high QoS degradation beyond the threshold. Note that, in this scenario, although {\small{Pred-Alloc}} leads to a high bandwidth allocation of 77.7\%, it keeps the mean QoS degradation $\beta_{\textit{max}(T)}$ under 1\%.
{\small{WC-CPO}} and {\small{WCSAC}} lead to a mean bandwidth allocation of 53.9\% and 56\%, and a mean QoS degradation of 8.98\% and 7.19\%, respectively. Although resulting in better generalization than the previous approaches, these results show that {\small{WC-CPO}} overfits to the worse-case traffic scenarios due to the exponential cost, while {\small{WCSAC}} adapts well to both average and worse-case scenarios.
\begin{figure*}[t]
\centering
\begin{minipage}[t]{0.66\columnwidth}
\centering
\includegraphics[width=\linewidth]{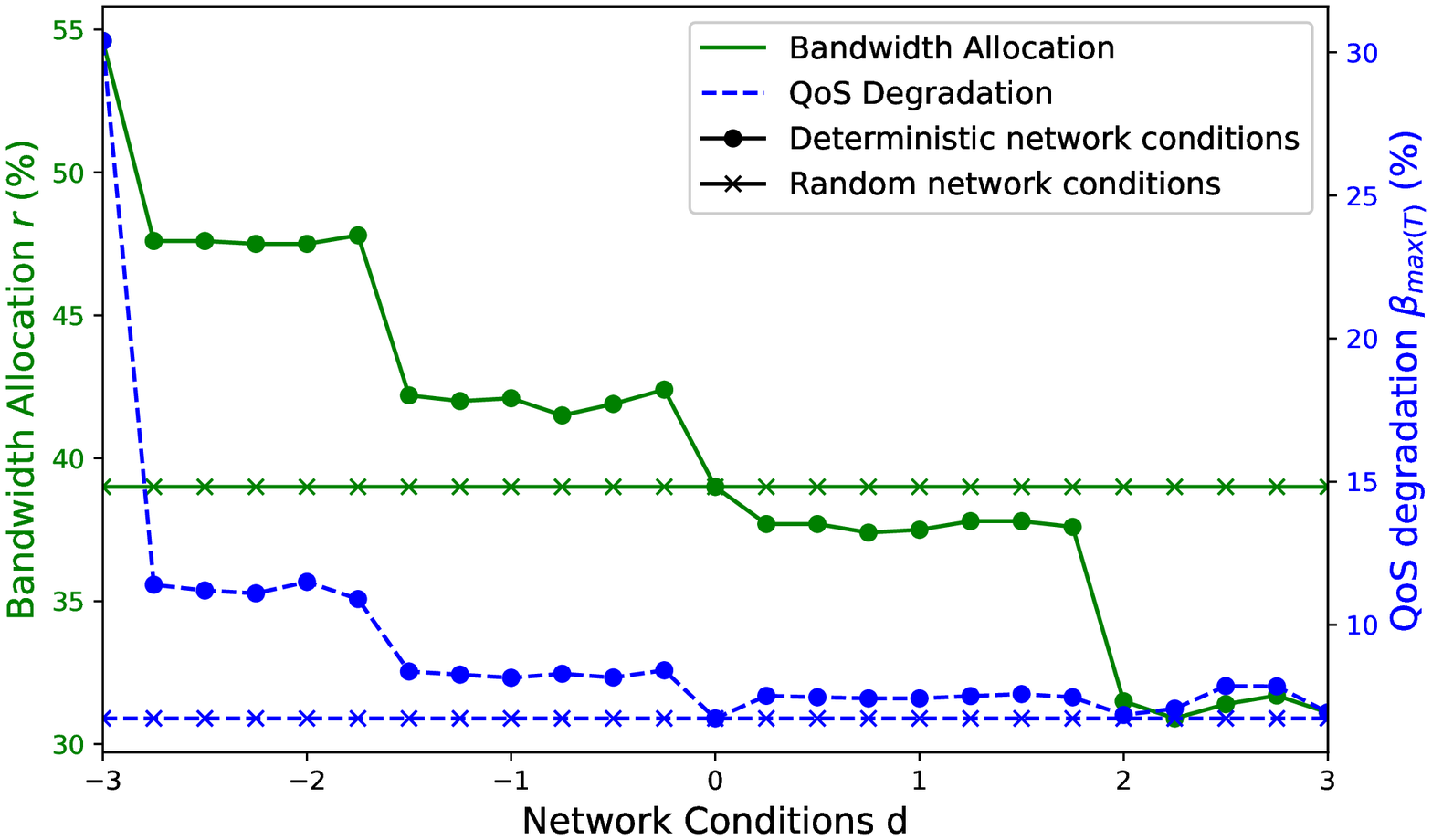}
\caption{WCSAC performance at different network conditions ($\beta_\textit{thresh}=10\%$)}
\label{net_conds}
\end{minipage}
\begin{minipage}[t]{0.66\columnwidth}
\centering
\includegraphics[width=\linewidth]{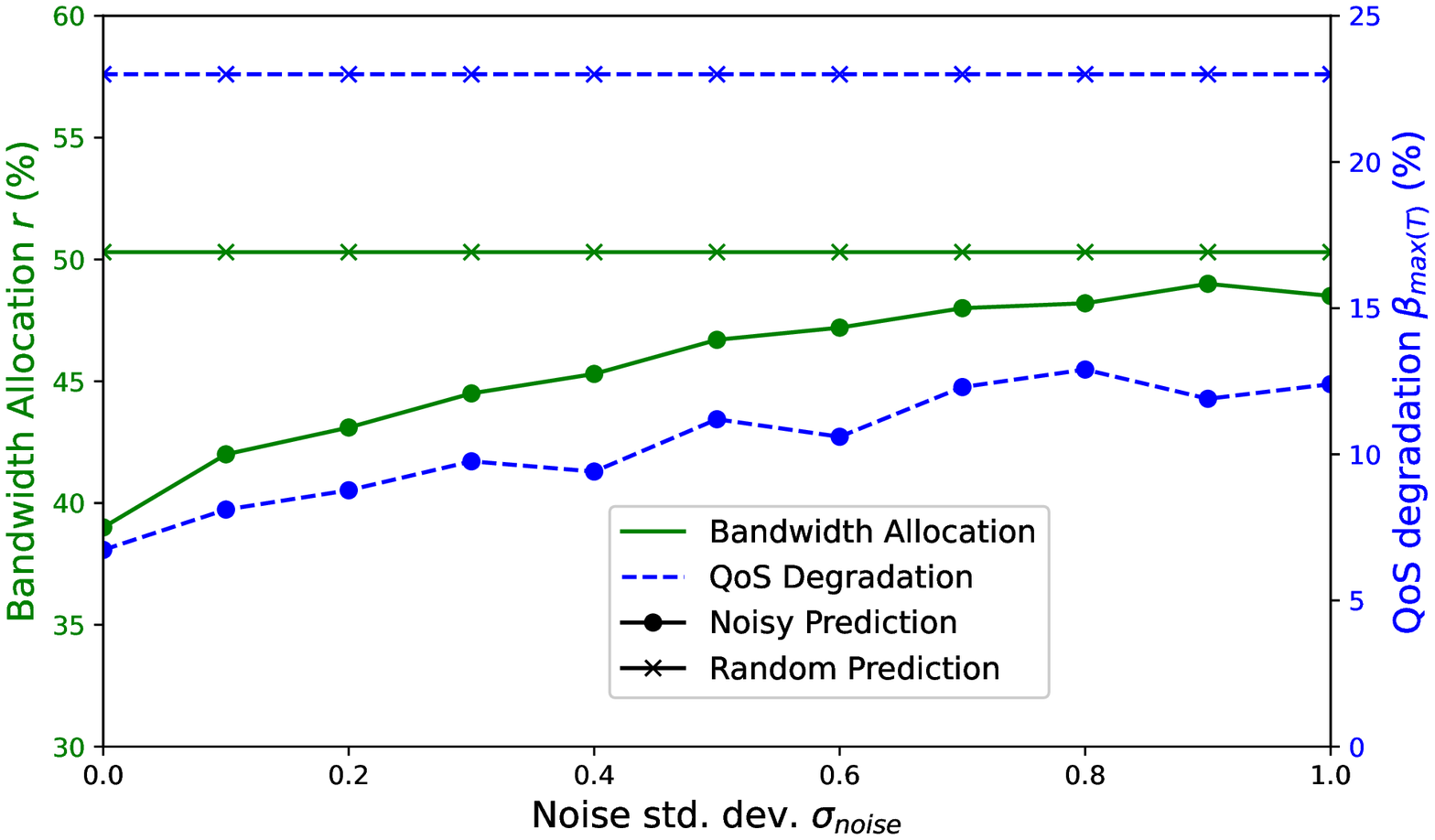}
\caption{WCSAC performance with inaccurate traffic prediction ($\beta_\textit{thresh}=10\%$)}
\label{noise}
\end{minipage}
\begin{minipage}[t]{0.66\columnwidth}
\centering
\includegraphics[width=\linewidth]{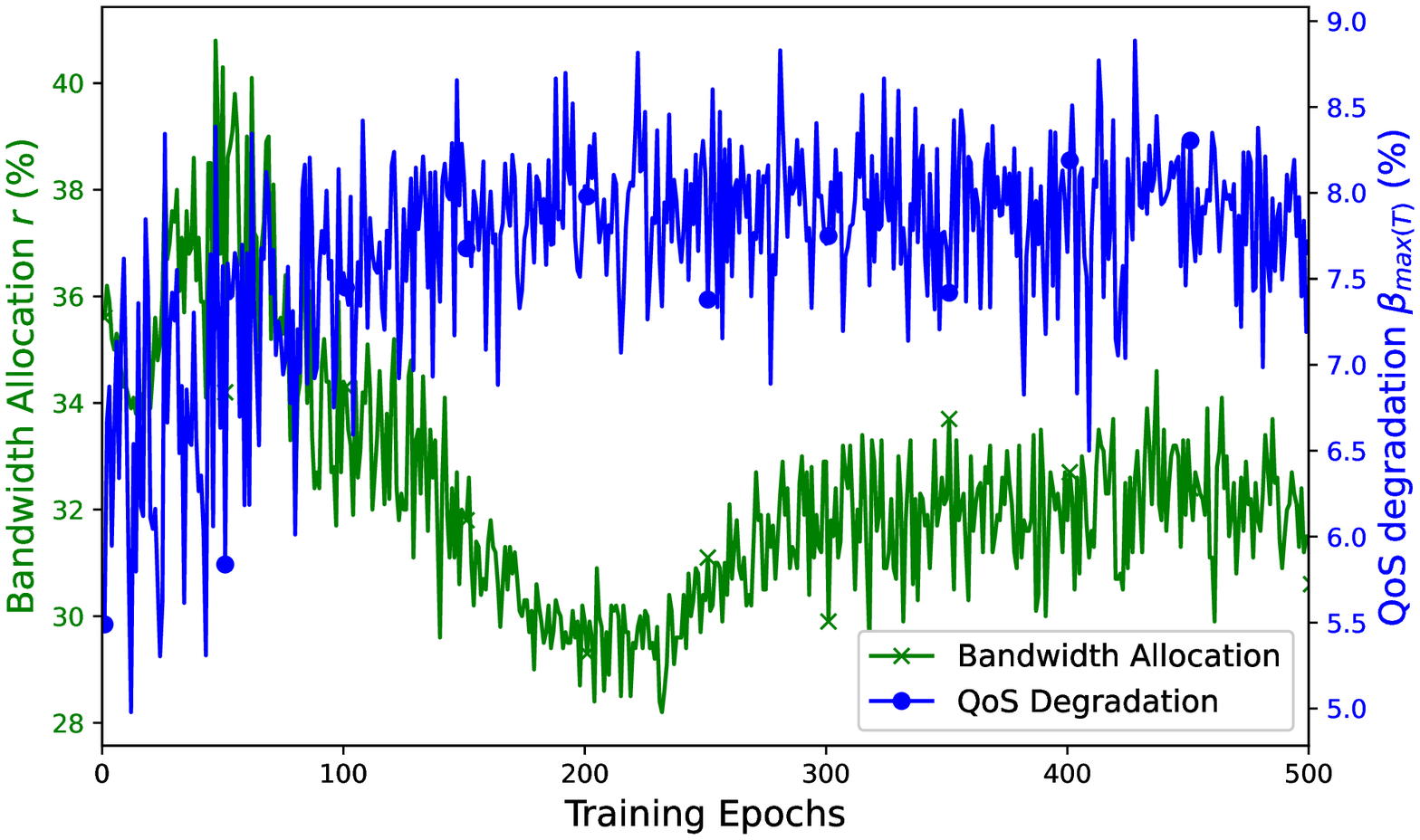}
\caption{WCSAC avg. BW and QoS degradation during fine-tuning on deterministic network condition $q^{det}_t(+1)$ with $\beta_\textit{thresh}=10\%$}
\label{finetune}
\end{minipage}
\vspace{-10pt}
\end{figure*}

\subsection{Robustness to Varying Network Conditions and Inaccurate Traffic Predictions}

First, we evaluate the proposed approach, on the dataset traffic, under different network conditions. To  control the severity of network conditions (\ie from favorable to worst-case), we assume them to be deterministic. In this case, $q^{det}_t(-3)$ and $q^{det}_t(+3)$ represent worst-case and best-case network conditions, respectively. 
In \fig{net_conds}, \blue{the horizontal lines show the bandwidth allocation and  QoS degradation when the QoS is randomly sampled using the network model output}, while the curves show their trend under static network conditions with varying $d$. \blue{Evidently, as the network conditions aggravate, both the QoS degradation and bandwidth allocation increase. The QoS degradation stays within the threshold $\beta_{\textit{thresh}}$ as long as $d \geq -1.5$. For $1.5 < d < 3$, the QoS degradation is slightly higher than the  threshold.
Finally, at $d = -3$, the QoS degradation hits 30.4\%. 
This is expected since \small WCSAC} constrains only the CVaR of the cost distribution at a given risk level (\ie $\alpha$) under the limit. On the other hand, as network conditions become more favorable, it results in an increase in resource efficiency and a decrease in QoS degradation.


Subsequently, we evaluate the sensitivity of the proposed approach to inaccurate traffic prediction. For this purpose, we use dataset traffic but introduce truncated Gaussian noise $\mathcal N(0, \sigma_{noise}^2)$ to the traffic probability distribution, at each step. After normalization, this resulting CDF is fed as the state to the RL agent. In Fig.~\ref{noise}, the horizontal lines show the resource allocation and the QoS degradation when the traffic prediction is fully random (\ie $\hat{x}_{tn}\sim\mathcal{U}(1,5)$), while the curves show these metrics as $\sigma_{noise}$ increases. We can observe that when prediction inaccuracy increases (\ie for higher $\sigma_{noise}$), both the QoS degradation and bandwidth allocation increase. However, even with fully randomized traffic prediction, the proposed approach shows a QoS degradation that is only 13\% higher than the threshold $\beta_{\textit{thresh}}$. This is attributed to the RL agent taking the current QoS degradation level into account when allocating resources, which acts as a feedback mechanism and allows the agent to allocate higher resources even if the traffic prediction is inaccurate.

\subsection{Fine-tuning Performance}


In a production network, the online data regarding a slice's traffic pattern and the network condition can be used to continuously improve the network model and resource scaling algorithm with fine-tuning. 
Since the traffic pattern and network conditions do not change during testing, learning to adapt to other scenarios is not required during fine-tuning. Therefore, we train the {\small WCSAC} algorithm with $\alpha=0.99$ (\ie risk-neutral), and at a lowered learning rate for 500 epochs. Fig.~\ref{finetune} shows the bandwidth allocation and the QoS degradation when {\small WCSAC} is fine-tuned on the dataset traffic with a favorable and deterministic network condition, \ie $q^{det}_t(+1)$. The optimal bandwidth allocation is achieved within 250 epochs while maintaining QoS degradation below the set threshold of 10\%. We evaluate the fine-tuned model using the best checkpoint during training. Evaluation results on the dataset traffic show a mean bandwidth allocation of 29.7\% and a mean QoS degradation of 7.85\%. Compared to previous results in \fig{net_conds}, there is 7.8\% less bandwidth usage while maintaining QoS degradation under the threshold.  
We also test this fine-tuned algorithm on the dataset traffic offset by 2 users, with randomly sampled network conditions, which leads to a mean bandwidth allocation of 56.9\% and a mean QoS degradation of 9.06\%. Comparing these to the results in \tab{table:ev}, we can conclude that the gain in performance during fine-tuning comes at only a slight cost to the generalization ability of the algorithm.

\section{Conclusion}\label{sec:conclusion}

In this work, we developed a novel framework utilizing a risk-constraint DRL algorithm, a regression-based network model and a traffic prediction module, for dynamically scaling slice resources in a 5G network. 
We proposed a regression-based network model to learn the distribution of QoS at any resource allocation and traffic, and utilized that to train the RL agent offline. To achieve generalization, the RL agent is fed with future traffic and current QoS degradation level, and is trained on randomized traffic. 
Our results show that the resulting RL agent is able to show similar performance as RL agents trained on the exact test-time traffic. Furthermore, it is able to maintain its performance across different real-world traffic patterns. Additionally, we demonstrated the agent's robustness under extreme network conditions and inaccurate traffic prediction. Finally, we showed that fine-tuning can be used to improve the performance further when the network condition or the slice traffic pattern is known in advance. 
As a future direction for this work, we intend to extend the evaluations over multiple types of resources and slices. We also plan to validate the algorithm on an expansive 5G testbed.


\section*{Acknowledgement}
This work was supported in part by Rogers Communications Canada Inc. and in part by a Mitacs Accelerate Grant.
\newpage
\bibliographystyle{IEEEtranN} 
\small{\bibliography{references}}

\end{document}